\documentclass[12pt]{iopart}
\usepackage{graphicx}
\usepackage{color}
\usepackage{iopams}

\newcommand{\dd}  { {\textrm d}}

\begin{document}

\title{Is Strangeness Still Strange at the LHC?}

\author{G G Barnaf\"oldi$^{1,2}$, G Fai$^1$, P L\'evai$^2$, B A Cole$^3$, and G Papp$^4$}
\ead{bgergely@rmki.kfki.hu}
\address{$^1$CNR, Department of Physics, Kent State University, Kent, OH 44242, USA}
\address{$^2$MTA KFKI RMKI, P.O. Box 49, Budapest 1525, Hungary}
\address{$^3$Nevis Laboratory, Columbia University, New York, NY 10027, USA}
\address{$^4$E\"otv\"os University, P\'azm\'any P. 1/A, Budapest 1117, Hungary}

\begin{abstract}
Strangeness production is calculated in a pQCD-based model (including 
nuclear effects) in the high transverse momentum sector, where pQCD is 
expected to work well. We investigate pion, kaon, proton and lambda 
production in $pp$ and heavy-ion collisions. Parton energy loss
in $AA$ collisions is taken into account. We compare strange-to-non-strange
meson and baryon ratios to data at RHIC, and make predictions for the LHC.
We find that these ratios significantly deviate from unity not only at RHIC
but also at the LHC, indicating the special role of strangeness at both
energies. 
\end{abstract}

\section{Introduction}

Strange particles have played a major role in high-energy physics since
the early 1950s. Originally, ``strange'' meant `created in pairs, with a
new quantum number, $s = \pm 1$, respectively'. It should also be
understood that the meaning of ``high-energy physics'' continued to
change in step with the construction of more and more powerful accelerators. 
In the collider era, the title question attracts renewed interest in 
connection with the extended volumes of high energy-density matter created 
in nucleus-nucleus ($AA$) collisions as the high-energy frontier moves 
from the Relativistic Heavy Ion Collider (RHIC) to the Large Hadron Collider 
(LHC). RHIC collides gold nuclei at 200 $A$GeV center-of-mass energy. 
The LHC will operate at more than an order of magnitude higher energies.

The historical roots stretch back to the 1940s: curious V-shaped tracks
were found in emulsion, primarily in $\pi \to \mu$ decays. The mysterious
'{\it V-particles}' were the first observed examples of matter with
strangeness content. Later the $ \Lambda $ particle was identified and
described by the additive quark model and the name ``strangeness''
was suggested by M. Gell-Mann in the middle of the '50s~\cite{GellMann}.

Another period of excitement was ushered in by the discovery of the $J/\Psi$ 
in the famous ``November Revolution'' of 1974~\cite{Jpsi}. This opened a new 
era of high-energy physics: strange quarks no longer represented the 
heaviest flavor, not present as valence contributions in everyday baryons.
It was expected that at high energies the new (``charm'') quarks will
play a role similar to that of the strange quarks at lower energies as
the small mass difference of the up, down, and strange quarks becomes negligible.
As a consequence, it is natural to expect that in the high energy-density
phase (quark-gluon plasma, QGP) up, down, and strange quarks have 
similar populations and play equivalent roles. This expectation is not 
completely borne out at RHIC; does it become a good approximation at 
LHC energies?  
 
The non-Abelian quark-gluon matter can not be observed directly,
but the abundances of final-state hadrons should be sensitive to the 
strangeness content. This is  the basis of the so-called 
``strangeness signature'' of the QGP, specifically the expected enhancement 
of strange anti-baryon production, as proposed by 
M\"uller and Rafelski~\cite{Muller} and independently by
B\'\i r\'o and Zim\'anyi~\cite{Biro}. 

What can we expect from the $\sqrt{s}=5.5$ $A$TeV $PbPb$ collisions,
where more final-state strange particles are produced with larger transverse 
momenta (``hard probes''), and perturbative quantum chromo-dynamics (pQCD) is 
to provide more precise predictions? It is interesting to ask whether 
the role of strangeness will finally change at these energies, where we are 
going to have a more precise ``microscope'' in the form of the LHC experiments.    
Do the $s$ quarks behave as light quarks ($u$ and $d$) and does
charm appear as the ``first massive'' flavor? 

In this paper we present calculations for $K/\pi$ and $ \Lambda /p $ ratios 
from $AA$ collisions at RHIC and LHC energies. We include the effect of
nuclear modifications on the produced particles. We present the results in terms
of double ratios of nuclear modification factors.


\section{Theoretical Model}

The particle ratios were calculated using hadron spectra as provided by 
our perturbative QCD improved parton model~\cite{Yi02}. The model is based 
on the factorization theorem 
and generates the invariant cross section as a convolution of (nuclear) 
parton distribution functions $f_{a/A}$, perturbative QCD cross sections 
$\dd \sigma^{ab \rightarrow cd }/ \dd  \hat t$, fragmentation functions 
$D_{\pi/c}$, and nuclear thickness functions $t_A$. 
We perform the calculation following
Refs.~\cite{Yi02,Levai0306,Levai0611,bggqm04,bggqm05,Bp02}:
\begin{eqnarray}
\label{hadX} 
\hspace{-1.0truecm} 
E_{h} \frac{\dd \sigma_{h}^{AA'}}{\dd ^3p_{h} } & \sim &
t_A(r) \,\, t_{A'}(|{\bf b} - {\bf r}|) \otimes 
f_{a/A}(x_a,Q^2;{\bf k}_{Ta}) \otimes  
f_{b/A'}(x_b,Q^2;{\bf k}_{Tb}) \otimes \nonumber \\
&\otimes & \frac{ \dd \sigma^{ab \rightarrow cd }}{\dd  \hat t }  
\otimes \frac{ D_{ h/c}(z_c,{\widehat Q}^2)}{\pi z_c^2} \,\, ,
\end{eqnarray}
where $Q^2$ and ${\widehat Q}^2$ represent the factorization and
fragmentation scales, respectively, $x_a$, $x_b$, and $z_c$ are momentum 
fractions, and ${\bf k}_T$-s stand for two-dimensional transverse momentum 
vectors. The initial state effects of shadowing and multiscattering are 
included following the ideas in Refs.~\cite{Yi02,Levai0306,Levai0611}.
The collision geometry is described by the impact parameter ($b$) dependent
Glauber nuclear thickness functions, $t_A(b)$.

For the nuclear parton distributions (PDFs) we have applied the MRST 
central gluon (cg) set~\cite{MRST} with the updated HIJING 
shadowing~\cite{HIJING,Shad_HIJ}.
Intrinsic transverse momentum and multiple scattering are treated according to
Ref.~\cite{Yi02}. For the fragmentation functions (FFs), we used a recent set by 
Albino, Kniehl, and Kramer~\cite{AKK1} to calculate the $K$, $\pi$, $p$ and 
$\Lambda$ spectra. All calculated hadron spectra are charge averaged, thus pions, 
kaons, protons, and lambdas are $\pi=(\pi^++\pi^-)/2$, 
$K=(K^++K^-)/2$, $p=(p^++p^-)/2$, and 
$\Lambda=(\Lambda+\bar{\Lambda})/2$, respectively.

In the model the energy loss of partons propagating through the high 
energy-density matter was taken into account along the lines of the 
GLV treatment~\cite{glv}. In this picture the amount of jet energy 
loss is parameterized in terms of the nuclear opacity $L/\lambda$, where 
$L$ is the average distance traveled by the parton in the medium and $\lambda$
stands for the mean free path. Opacity $L/\lambda = 0$ signifies no jet
quenching, high values correspond to central collisions of heavy nuclei.  

We are interested in the nuclear modification for strange species relative to 
hadrons not containing strange quarks, i.e. pions and protons.
For this purpose we introduce the double ratio of different hadron's
nuclear modification factors,
\begin{equation}
\hspace{-2.0truecm}R^{s\bar{s}}_{AA}(p_T)=\frac{R_{AA}^{h^s}(p_T)}{R_{AA}^{h}(p_T)} = 
\frac{E_{h^s} \dd \sigma^{AA}_{h^s}/\dd^3 p_{h^s}}
{E_{h^s} \dd \sigma^{pp}_{h^s}/\dd^3 p_{h^s}} \left/ \/
\frac{E_{h} \dd \sigma^{AA}_{h}/\dd^3 p_{h}}
{E_{h} \dd \sigma^{pp}_{h}/\dd^3 p_{h}} \right.
= \left[ \frac{h^s}{h} \right]_{AA} \left/  \left[ \frac{h^s}{h}\right]_{pp}, \right.
\label{rss_def}
\end{equation}
where we used the notation $h^s$ for charged-averaged strange hadrons and 
$h$ for non-strange hadrons. The nuclear modification factor is 
defined as 
\begin{equation}
\hspace{-2.0truecm}R^{h}_{AA}(p_T)=\frac{1}{\langle N_{bin} \rangle } \cdot
\frac{E_{h} \dd \sigma^{AA}_{h}/\dd^3 p_{h}}
{E_{h} \dd \sigma^{pp}_{h}/\dd^3 p_{h}}  \,\, .
\label{raa_def}
\end{equation}
Here $\langle N_{bin} \rangle$ is the average number of binary
collisions in the various impact-parameter bins, which is 
determined by the geometry of the collisions and cancels from the double ratio
(\ref{rss_def}).

The last equation of (\ref{rss_def}) indicates that the double ratio of 
nuclear modification factors can be obtained as the double ratio of 
strange-to-non-strange ratios in $AA$ to $pp$ collisions. 

\section{Strange Particle Ratios at RHIC Energies}

\begin{figure}
\begin{center}
\includegraphics[width=7.5truecm,height=8.0truecm]{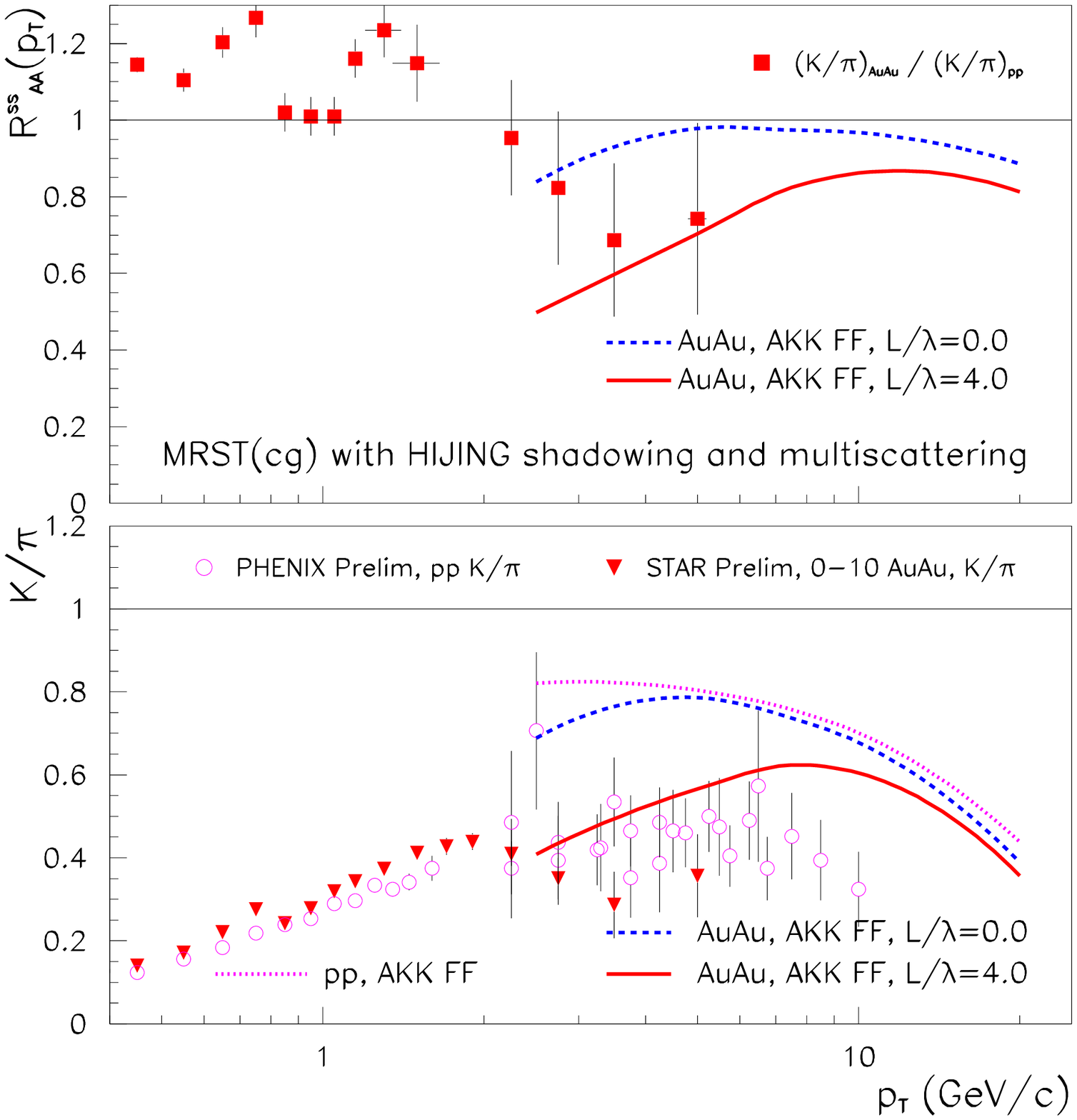}
\includegraphics[width=7.5truecm,height=8.0truecm]{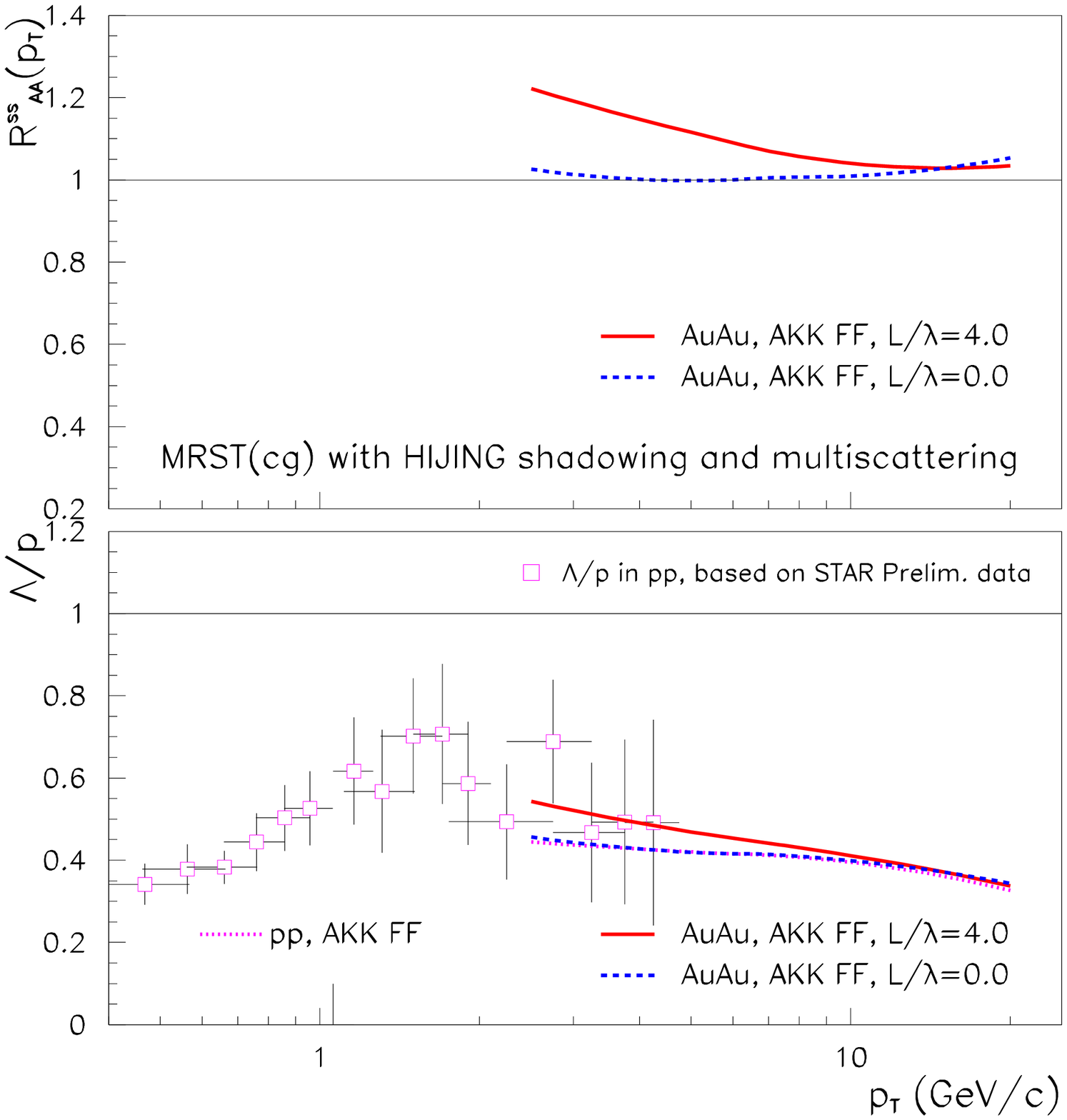}
\end{center}
\vspace*{-1.0truecm}
\caption{Calculated $K/\pi$ ratios in $AuAu$ and $pp$ 
collisions compared to PHENIX $pp$~\cite{PHENIX}
STAR $AuAu$~\cite{STAR_data,STAR_data2} data  
({\sl left side}) and $\Lambda/p$ ratios  
with $pp$ collision data from STAR~\cite{Heinz_STAR,STAR_proton} ({\sl right side}).
The bottom panels display the strange-to-non-strange ratios, 
while the top panels show the double ratios defined by eq.~(\ref{rss_def}).}
\label{fig1}
\end{figure}

On the {\sl left side} of Fig. \ref{fig1}, in the bottom panel,
$K/\pi$ ratios are plotted as measured by the PHENIX collaboration in $pp$
collisions~\cite{PHENIX} and by the STAR collaboration in $AuAu$ 
collisions~\cite{STAR_data,STAR_data2} at $\sqrt{s}= 200$ $A$GeV RHIC energy, 
together with calculated $K/\pi$ ratios 
in $pp$ collisions ({\sl dotted line}) and $AuAu$ collisions without jet 
energy loss ($L/\lambda=0$, {\sl dashed}) and with opacity $L/\lambda=4$ ({\sl solid
line}) for the  most central ($0-10\%$) $AuAu$ collisions. In the top panel
we show the calculated double ratios (\ref{rss_def}) relative to $pp$ collisions 
without ({\sl dashed}) and with ({\sl solid line}) jet energy loss, compared to those 
obtained from the data. (Note that the experimental double ratios naturally have
large uncertainties.) 
 
The experimental $K/{\pi}$ ratios are very similar to each other
at low $p_T$, although a slight hint of 'strangness enhancement'
is indicated by the $\approx$ 20 \%  increase in nucleus-nucleus 
collision relative to proton-proton collisions. However,
this region can not be investigated by pQCD calculations.
At high $p_T$, forgeting the large error bars for a moment,
the data in the top left panel show some suppression of kaons in 
$AuAu$ collisions. Do the theoretical calculations
indicate a similar tendency? The answer is yes, as displayed by Fig.~1.
Introducing stronger jet quenching with larger opacity in $AuAu$ 
collisions, the $K/\pi$ ratio is decreasing. At very high $p_T$ the 
data appear to fall faster than the theoretical curves. More 
data are needed with higher precision for a final answer. One needs to also 
keep in mind that data from PHENIX~\cite{PHENIX} and STAR~\cite{STAR_data,STAR_data2} 
were combined to obtain the experimental points in the top panel. 

Considering the strange-to-non-starnge ratio,
while the proton-proton calculations overestimate the data by up to a factor 2,
there is a hint of potential agreement between the data and the calculations
with an opacity of $L/\lambda=4$ in the bottom panel. Unfortunately, as 
mentioned above, the calculations can not be extended to lower momenta, 
where pQCD is no longer reliable (technically due to PDF and FF limitations). 

On the {\sl right side} of Fig. \ref{fig1} we show similar information for 
$\Lambda/p$ ratios. Here, the calculated curves run very close to each other
in the bottom panel. The data are for $pp$ collisions from 
Refs.~\cite{Heinz_STAR,STAR_proton}. (We are not aware
of measured $AuAu$ $\Lambda/p$ ratios, and we encourage such analysis.) 
In summary, the mesonic $K/\pi$ ratio is more sensitive to jet energy loss
in heavy-ion collisions than the $\Lambda/p$ baryonic ratio.
The reason is very simple: at lower $p_T$ meson production is dominated 
by gluon fragmentation, which slowly turns into the dominance of 
quark fragmentation with increasing $p_T$. On the other hand, baryon production 
is dominated by leading quark fragmentation in a wide $p_T$ region.
In this sense the investigation of ${\overline \Lambda}/{\overline p}$
may be more interesting because the gluons have a larger contribution
to this ratio. 

\section{Is Strangeness Still Strange at LHC Energies? }

\begin{figure}
\begin{center}
\includegraphics[width=7.5truecm,height=8.0truecm]{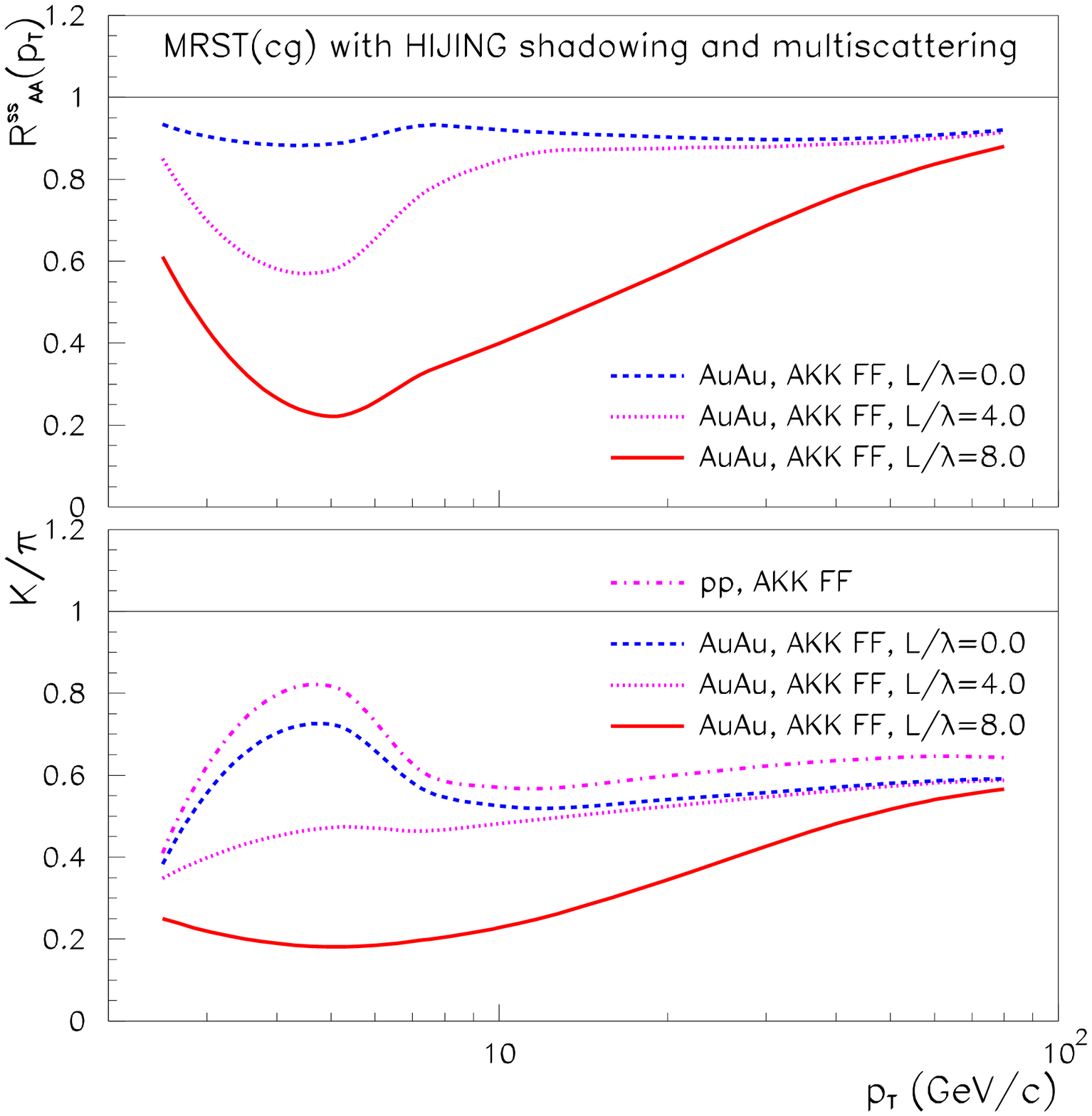}
\includegraphics[width=7.5truecm,height=8.0truecm]{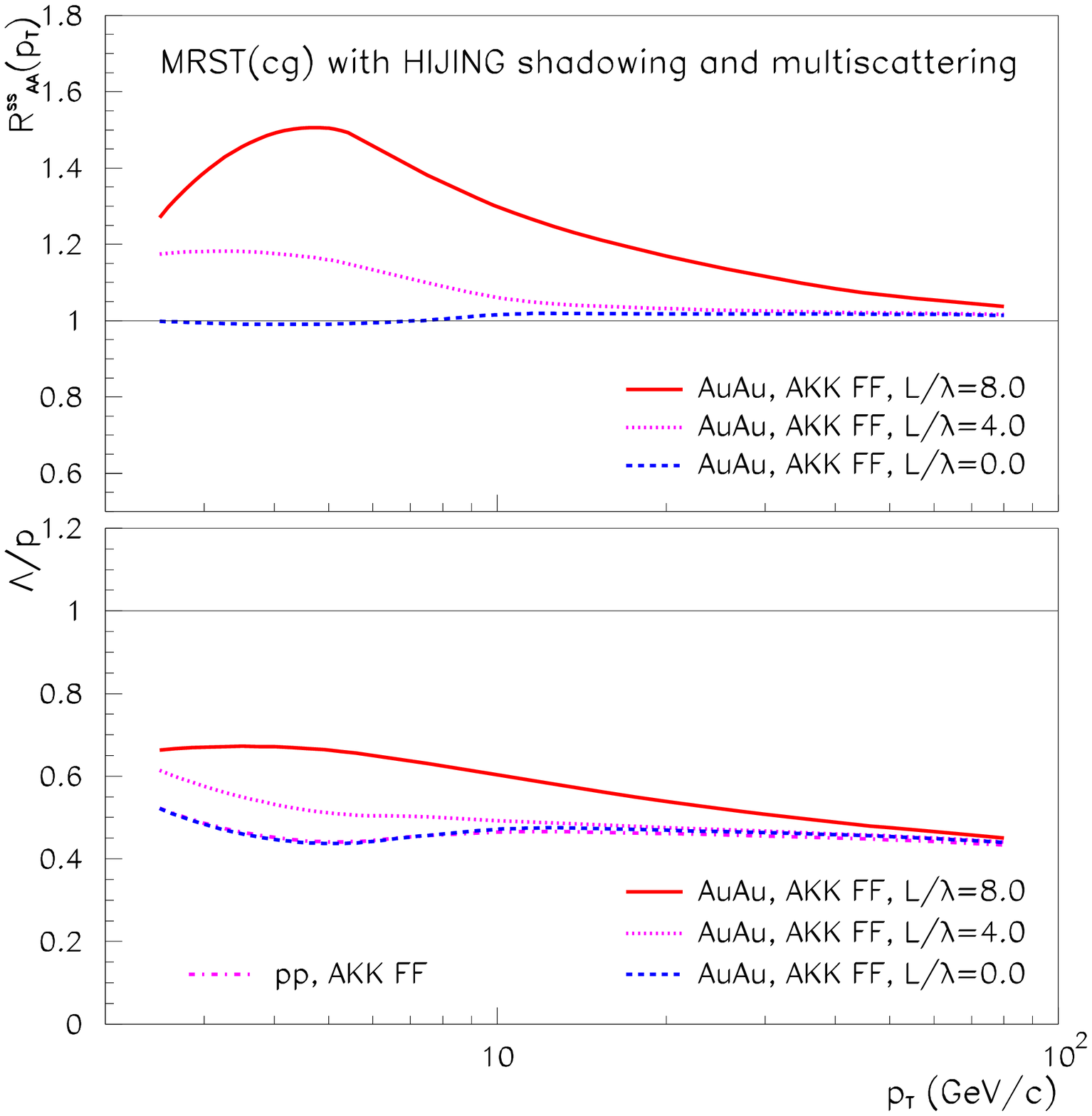}
\end{center}
\vspace*{-1.0truecm}
\caption{Calculated  $K/\pi$ ({\sl left side}) and $\Lambda/p$ 
({\sl right side}) ratios (bottom) and double ratios (top) in $PbPb$ 
collisions at $\sqrt{s}=$ 5.5 $A$TeV with different opacities.}
\label{fig2}
\end{figure}

We repeat our calculations for $PbPb$ collisions for $\sqrt{s}= 5.5$ 
$A$TeV LHC energy. The results are displayed on Fig.~\ref{fig2}.
The opacity is expected to be higher at LHC due to the higher available
energy. Using a simple $\dd N / \dd y \sim 1500-3000$ estimate, 
we obtain $L/\lambda \approx 8$ in the most central $0-10\%$ $PbPb$ 
collisions.  For comparison, we also plot the results with
$L/\lambda = 0$ and $4$. On the bottom panel in the {\sl left side}  
we show the $K/\pi$ ratios with the double ratios on top, up to
high transverse momenta. The {\sl right side} contains the prediction
for the $\Lambda/p$ ratios. 

Results at LHC energies are somewhat similar to those at RHIC energies. 
The lower- and intermediate-$p_T$ variation of the hadron ratios 
arises from the different strengths of the jet quenching for quark and gluon 
contributions~\cite{KaonJet,BGGLHC}. Due to the quark dominated fragmentation, 
the difference disappears at high-$p_T$ in the ratios.

\section{Conclusions}

We find that strangeness still behaves differently from the up and down
quark contributions at LHC energies in that the $K/\pi$ and $\Lambda/p$ ratios
are still below unity at $\sqrt{s}=$ 5.5 $A$TeV. The strange-to-non-strange
ratios are similar for mesons and baryons, and are expected to be similar
at LHC to what was seen at RHIC. On the other hand, mesonic ratios show
more structure in the intermediate momentum region, because gluonic
contributions have a bigger role in this momentum window.

\section*{Acknowledgments}

Special thanks to Prof. John J. Portman for computer
time at Kent State University. Our work was supported in part by
Hungarian OTKA T047050, NK62044, and IN71374, by the U.S. Department
of Energy under grant U.S. DE-FG02-86ER40251, and jointly by the U.S.
and Hungary under MTA-NSF-OTKA OISE-0435701.


\end{document}